\documentstyle[twocolumn,epsf,subfigure]{jpsj}

\def\runtitle{}
\def\runauthor{Ken {\sc Yokoyama}} 

\title{
Crossover from Fermi to Non-Fermi Liquid
in Two-Dimensional\\
Interacting Fermions
}

\author{
Ken {\sc Yokoyama}\footnote{E-mail: yokoyama@watson.phys.s.u-tokyo.ac.jp}
}

\inst{
Department of Physics, Tokyo University, Tokyo 113
}

\recdate{June 5, 1998}

\abst{
Self-energy at zero temperature is investigated up to the third-order
of interaction using
one-patch model in two dimensions, whose interaction process
corresponds to $g_4$-process of $g$-ology model in one dimension.
The self-energy $\Sigma^R({\mib k},\epsilon)$ diverges at $\epsilon=\xi_k$,
and the contribution from the particle-hole process
in third-order self-energy diagrams has a stronger divergence
compared to the one from the particle-particle process. This
implies that the loop-cancellation in the forward scattering is
insufficient due to the effect of the warping of the Fermi surface.
The strong energy dependence of the self-energy
in the vicinity of $\epsilon=\xi_k$
implies the existence of the crossover from Fermi to non-Fermi liquid
behavior as the momentum becomes away from the Fermi momentum,
and this crossover is enhanced as interaction becomes stronger.
}

\kword{
two dimensional electron gas, forward scattering, crossover from Fermi 
to non-Fermi liquid, loop-cancellation
}

\begin{document}
\sloppy
\maketitle

\newcommand{\rRe}{{\rm Re}\,}
\newcommand{\iIm}{{\rm Im}\,}
\newcommand{\sign}{{\rm sgn}}
\newcommand{\bm}{\bf\mit}

The low-energy excitation of the interacting fermions 
with short-range force is established as
Tomonaga-Luttinger (TL) liquid in one dimension and Fermi liquid in three
dimensions, while as for two dimensions it is still in controversy.
In case of one dimension, forward scattering processes,
which are denoted $g_2$- and $g_4$-process in $g$-ology model,
leads to TL liquid.\cite{bib:Voitrev}
The main features of TL liquid are the following two
things: vanishing jump of momentum distribution at Fermi momentum
and spin-charge separation. The $g_2$-process is
related to the former, and the $g_4$-process
to the latter in the following sense.
Taking account of only $g_{2\parallel}$- or $g_{2\perp}$-process,
the velocity of spin and charge excitations, $v_\rho$ and $v_\sigma$,
respectively, become different, which indicates the existence of
spin-charge separation and results in the two-peak
structure of spectral-weight.\cite{bib:Meden,bib:Voit}
On the other hand,
if we consider only $g_{4\perp}$-process (here
$g_{4\parallel}$-process related term cancels if we neglect momentum
dependence of the coupling constant),
the parameters $K_\rho$ and $K_\sigma$, which equal to $1$ for
free fermions and characterize
anomalous power-laws of various correlation functions, deviate
from $1$. This also leads to vanishing jump of momentum distribution
at Fermi surface. 

In case of two dimensions, it was suggested
that anomalous behavior of forward scattering phase-shift
leads to non-Fermi liquid even at weak-coupling
quite similarly to
TL liquid.\cite{bib:Anderson1,bib:Anderson2}
But in this stage, there is no theory which confirms this
possibility. There is also no signal of non-Fermi liquid state
from many-body perturbation
approach in two dimensions.\cite{bib:Hodges,bib:Fab,bib:Stamp,bib:Met,bib:Metrev,bib:Engel,bib:FHN,bib:Ken,bib:com1,bib:com2,bib:com3,bib:rep}
In the following we
investigate this possibility from perturbation
theory in detail.

To start with, we consider
the following correspondence between the model in
one and two dimensions.
We obtain low-energy effective theories by integrating out degrees
of freedom of electrons far from Fermi points (or surface),
which is so-called the elimination of fast modes.\cite{bib:Metrev,bib:Shankar}
In one dimension, the low-energy effective theory is $g$-ology
model, and there are two branches corresponding to two Fermi points.
In case of two dimensions, the low-energy effective theory
has only degrees of freedom of electrons in a thin shell
with thickness $\Lambda$ around the Fermi surface.
Since momenta of electrons are allowed only within the thin shell,
interaction processes are extremely restricted; only three
kinds of processes shown in fig.\ \ref{fig:scat}, i.e.,
forward, exchange and Cooper scatterings,
are allowed (here we neglect Umklapp process).\cite{bib:Metrev,bib:Shankar}
Dividing the thin shell around the Fermi surface
to many small patches of the size $\Lambda\times\Lambda$,
we obtain the following low-energy effective action for zero temperature;
\begin{subeqnarray}
  {\cal S}\;\,\equiv\hspace{5mm}
    &\hspace{-1.5cm}\lefteqn{{\cal S}_0+{\cal S}_{forward}+{\cal S}_{exchange}
    +{\cal S}_{Cooper}}&
    \label{eq:model}\\
  {\cal S}_0&=&\hspace{-3mm}
    \sum_{\sigma}\int_{k\epsilon}
    Z_{k}^{-1}(i\epsilon-\xi_k)
    c^\dagger_{k,\sigma}c_{k,\sigma}
    \label{eq:H0}\\
  {\cal S}_{forward}&=&-\frac{1}{2}\sum_{ij}
    \sum_{\sigma\sigma^\prime}\int_{k\epsilon}\int_{k^\prime\epsilon^\prime}
    \int_{q\omega}\nonumber\\
  & &\hspace{-5mm}{g_F}^{\sigma\sigma^\prime}_{kk^\prime}(q)
    c^\dagger_{k+q,\sigma}
    c^\dagger_{k^\prime-q,\sigma^\prime}c_{k^\prime,\sigma^\prime}
    c_{k,\sigma}
    \label{eq:Hforward}\\
  {\cal S}_{exchange}&=&-\frac{1}{2}\sum_{i\ne j}
    \sum_{\sigma}\int_{k\epsilon}\int_{k^\prime\epsilon^\prime}
    \int_{q\omega}\nonumber\\
  & &\hspace{-5mm}{g_E}^{\sigma-\sigma}_{kk^\prime}(q)
    c^\dagger_{k+q,\sigma}
    c^\dagger_{k^\prime-q,-\sigma}
    c_{k,-\sigma}c_{k^\prime,\sigma}
    \label{eq:Hexchange}\\
  {\cal S}_{Cooper}&=&-\frac{1}{2}\sum_{i\ne j}
    \sum_{\sigma\sigma^\prime}\int_{k\epsilon}\int_{k^\prime\epsilon^\prime}
    \int_{q\omega}\nonumber\\
  & &\hspace{-5mm}{g_C}^{\sigma\sigma^\prime}_{kk^\prime}(q)
    c^\dagger_{k,\sigma}
    c^\dagger_{-k-q,\sigma^\prime}c_{-k^\prime+q,\sigma^\prime}.
    c_{k^\prime,\sigma}
    \label{eq:Hcooper}
\end{subeqnarray}
where
\begin{equation}
  \int_{k\epsilon}\equiv \int \frac{d^2k}{(2\pi)^2}
  \int_{-\infty}^{\infty} \frac{d\epsilon}{2\pi},
\end{equation}
and $c^\dagger_{k,\sigma}$ and $c_{k,\sigma}$ are Grassmann variables
for fermion with momentum and energy $k=({\mib k},i\epsilon)$
and spin $\sigma$,
and $g_F$, $g_E$ and $g_C$ are coupling constants for forward, exchange and
Cooper processes.
Denoting patches of the size $\Lambda\times\Lambda$ as $\Lambda_i$
($i$ is an index of patch),
the integration for momenta ${\mib k}$, ${\mib k^\prime}$ and ${\mib q}$ in 
eqs.\ (\ref{eq:Hforward}c), (\ref{eq:Hexchange}d) and (\ref{eq:Hcooper}e)
are performed in the region where
$k,k+q\in \Lambda_i$ and $k^\prime,k^\prime-q\in \Lambda_j$ are
satisfied.
${\cal S}_{Cooper}$ is a term related to Cooper instability in case of
attractive interaction. 
Regarding patches in two dimensions as analogs of branches in one dimension,
we can make correspondence from g-ology
model to the model given by the action of the form
eq.\ (\ref{eq:model}). Namely,
${\cal S}_{forward}$ corresponds to $g_2$- and $g_4$-terms,
and ${\cal S}_{exchange}$ to $g_1$-term in $g$-ology model.
As for ${\cal S}_{forward}$,
the case $i=j$ and $i\ne j$ correspond to
$g_2$- and $g_4$-terms, respectively.\cite{bib:Metrev,bib:Shankar}
\def\FIGSCALE{2.6cm}
\begin{figure}[t]
  \begin{center}
    \begin{tabular}[t]{c}
      \subfigure[]{\epsfxsize=\FIGSCALE\epsfbox{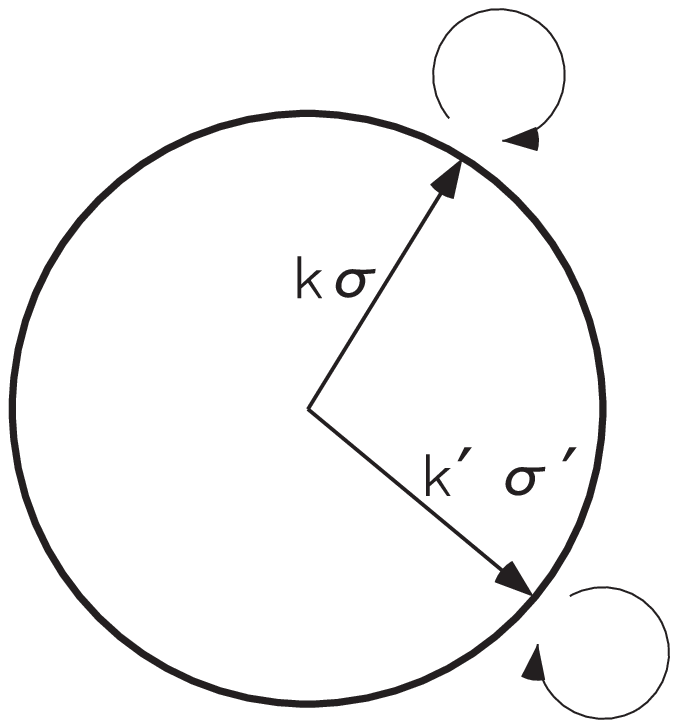}}
      \subfigure[]{\epsfxsize=\FIGSCALE\epsfbox{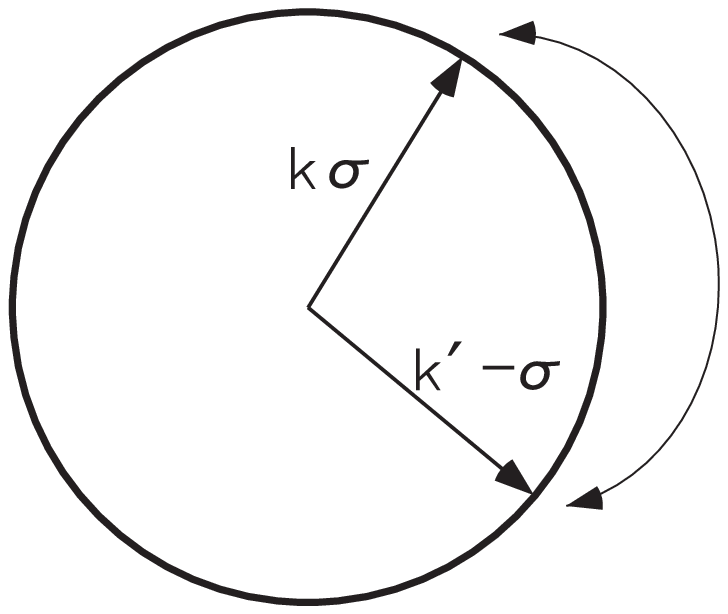}}
      \subfigure[]{\epsfxsize=\FIGSCALE\epsfbox{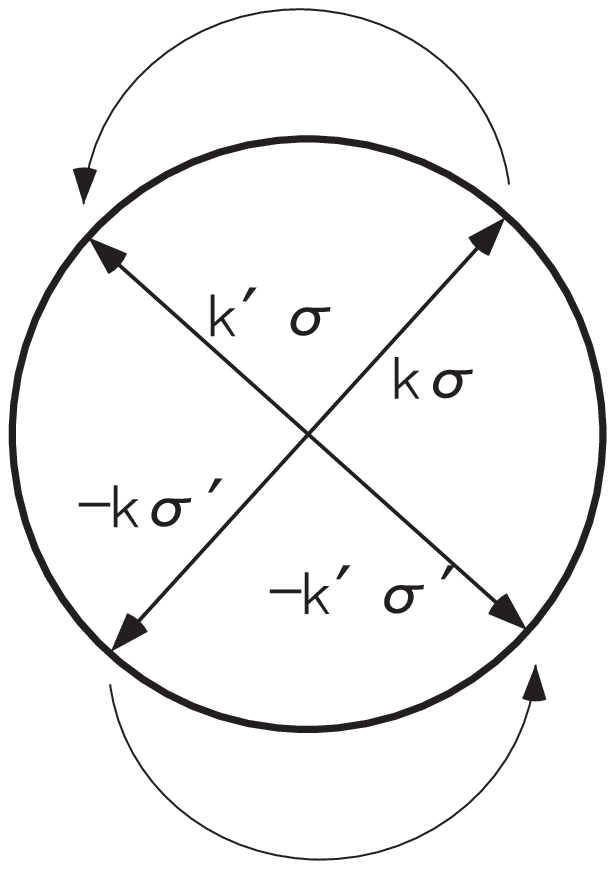}}
    \end{tabular}
  \end{center}
  \caption{Three kinds of interaction processes. (a)Forward scattering
        ($k\sigma,k^\prime\sigma^\prime\rightarrow
        k\sigma,k^\prime\sigma^\prime$).
        (b)Exchange scattering
        ($k\sigma,k^\prime-\sigma\rightarrow k^\prime-\sigma,k\sigma$).
        (c)Cooper scattering
        ($k\sigma,-k\sigma^\prime\rightarrow
        k^\prime\sigma,-k^\prime\sigma^\prime$).}
  \label{fig:scat}
\end{figure}

In one dimension, forward scatterings, i.e., $g_2$- and $g_4$-processes,
lead to TL liquid, and the question
is what character the low-energy excitations have in the
presence of the term ${\cal S}_{forward}$ in eq.\ (\ref{eq:model})
in two dimensions.
In the following, we consider the simplest case,
in which there exists only
$i=j$ term in ${\cal S}_{forward}$,
and calculate the self-energy up to the third order of interaction.
In one dimension, this process
leads to the spin-charge separation.
Following the procedure shown graphically in fig.\ \ref{fig:disp},
we introduce the model
described by an action of the form
${\cal S}={\cal S}_0+{\cal S}_I$, where
\begin{equation}
  {\cal S}_0=
    \sum_{\sigma}\int_{k\epsilon}
    Z^{-1}(i\epsilon-\xi_k)
    c^\dagger_{k,\sigma}c_{k,\sigma}
  \label{eq:model1a}
\end{equation}
and
\begin{eqnarray}
  {\cal S}_I&=&-\frac{U}{2}
    \sum_{\sigma}\int_{k\epsilon}\int_{k^\prime\epsilon^\prime}
    \int_{q\omega}\!\!
    c_{k+q,\sigma}^\dag c_{k^\prime-q,-\sigma}^\dag
    c_{k^\prime,-\sigma} c_{k,\sigma}.
  \label{eq:model1b}
\end{eqnarray}
Assuming ${g_F}^{\sigma-\sigma}_{kk^\prime}(q)$ is an analytic function
in the vicinity of $k=k^\prime$ and $q=0$,
and neglecting $k$, $k^\prime$ and $q$-dependences of
${g_F}^{\sigma-\sigma}_{kk^\prime}(q)$ in a patch,
we have replaced ${g_F}^{\sigma-\sigma}_{kk^\prime}(q)$ to a constant $U$
in eq.\ (\ref{eq:model1b}).
The renormalization factor $Z_k$ is also replaced to the constant $Z$
in eq.\ (\ref{eq:model1a}).
The origin of momentum and $k_x$ and $k_y$-axes are taken as shown
in fig.\ \ref{fig:disp}, and momentum cut-offs are introduced as
$|k_x|, |k_y| < \Lambda$.
We approximate the energy dispersion
as\cite{bib:Ogata}
\begin{equation}
  \xi_k = vk_x + \frac{A}{2}k_y^2.
  \label{eq:disp}
\end{equation}
Furthermore, we define cut-off energies ${\tilde \epsilon}_x$ and
${\tilde \epsilon}_y$ and the constant ${\tilde U}$ for later convenience as
\begin{equation}
  {\tilde \epsilon}_x\equiv v\Lambda, \;{\tilde \epsilon}_y\equiv
  A\Lambda^2, \;{\tilde U}\equiv U\Lambda^2
\end{equation}
Using this model,
we evaluate the contributions of the diagram shown in
figs.\ \ref{fig:3rd}(a) and \ref{fig:3rd}(b),
which we denote $\Sigma_{pp}^R({\mib k},\epsilon)$ and 
$\Sigma_{ph}^R({\mib k},\epsilon)$,
respectively, where ${\mib k}=(k,0)$.
\def\FIGSCALE{8cm}
\begin{figure}[t]
  \begin{center}
    \epsfxsize=\FIGSCALE\epsfbox{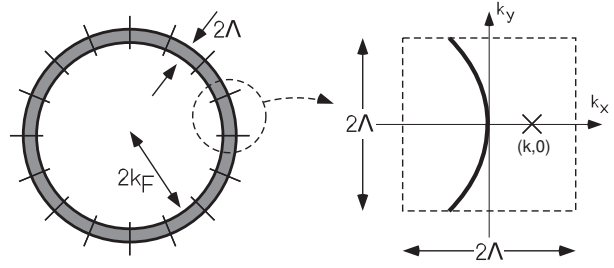}
  \end{center}
  \caption{Taking account of only the process corresponding to $g_4$-process
    in one dimension, we make a one-patch model.}
  \label{fig:disp}
\end{figure}

Firstly, we consider the contribution of the diagram in fig.\ \ref{fig:3rd}(a),
$\Sigma_{pp}^R({\mib k},\epsilon)$.
$\Sigma_{pp}({\mib k},\epsilon+i\delta)$ is expressed as
\begin{eqnarray}
  \lefteqn{\hspace{-4mm}\Sigma_{pp}^R({\mib k},\epsilon)=
    U^3\!\!\int_{q\omega}
    \left[
    \sign(\omega)\,\iIm
    \left[K^R({\mib q},\omega)\right]^2\,G_{q-k}^A(\omega-\epsilon)
    \right.}\nonumber\\
  & &\hspace{1cm}\left.+
    \sign(\omega-\epsilon)\,
    \left[K^R({\mib q},\omega)\right]^2\,
    \iIm G_{q-k}^R(\omega-\epsilon)\right].
\end{eqnarray}
Here $K^{R}({\mib q},\omega)$ is a particle-particle correlation function
defined as
\begin{equation}
  K^{R}({\mib q},\omega)\equiv\int_{kx}
    \sign(x)\,G_{q-k}^R(\omega-x)\,
    \iIm G_k^R(x),
\end{equation}
which is expressed approximately for $|\omega|, |vq_x|, |Aq_y^2|\ll
{\tilde \epsilon}_x, {\tilde \epsilon}_y$ as
\begin{equation}
  K^{R}({\mib q},\omega)\simeq\left\{\begin{array}{l}
    \displaystyle{K_0+\frac{iZ^2\omega}{4\pi vA^{1/2}
      (\omega-vq_x-Aq_y^2/4)^{1/2}}}\\[5mm]
    \hspace{2cm}\mbox{$(\omega-vq_x-Aq_y^2/4>0)$}\\[5mm]
    \displaystyle{K_0+\frac{Z^2\omega}{4\pi vA^{1/2}
      (-\omega+vq_x+Aq_y^2/4)^{1/2}}}\\[5mm]
    \hspace{2cm}\mbox{$(\omega-vq_x-Aq_y^2/4<0)$,}
  \end{array}\right.
  \label{eq:pp_cor}
\end{equation}
where
\begin{equation}
  K_0\equiv\lim_{q\rightarrow 2k_F}\lim_{w\rightarrow 0}K^R({\mib q},w)
     =\frac{Z^2\Lambda}{2\pi^2v}.
\end{equation}
Here the sequence of the limiting procedure is important reflecting the
singularity of the particle-particle correlation in the vicinity of
${\mib q}=2{\mib k}_F$.
We extract the term which contains singular part of self-energy
$\Sigma_{pp}^R({\mib k},\epsilon)$ in $k$ and $\epsilon$, which is
denoted as ${\Sigma_{pp}^R}^\prime({\mib k},\epsilon)$ and
defined as
\begin{eqnarray}
  \lefteqn{\hspace{-4mm}
    {\Sigma_{pp}^R}^\prime({\mib k},\epsilon)}\nonumber\\
  &&\hspace{-4mm}\equiv
    -U^3\!\!\int\!\!\frac{d^2q}{(2\pi)^2}\!\!
    \int_0^\epsilon\!\frac{d\omega}{\pi}
    \left[K^R({\mib q},\omega)\right]^2\,\iIm G_{q-k}^R(\omega-\epsilon).
    \label{eq:s_pp}
\end{eqnarray}
The analytic part of the self-energy, i.e., 
$\Sigma_{pp}^R({\mib k},\epsilon)-{\Sigma_{pp}^R}^\prime({\mib
  k},\epsilon)$, is considered to be related to various
renormalizations,
and whose effect can be absorbed into the renormalizations of
the constants $v$, $A$ and $Z$.
Substituting eq.\ (\ref{eq:pp_cor}) to eq.\ (\ref{eq:s_pp}), we obtain
\begin{eqnarray}
  \lefteqn{{\Sigma_{pp}^R}^\prime({\mib k},\epsilon)}\nonumber\\
  & &\hspace{5mm}=\left\{\begin{array}{ll}
      \displaystyle{c_1\epsilon+\frac{iZ^5{\tilde U}^3{\tilde K}_0}
	{8\pi^3{\tilde \epsilon}_x^2{\tilde \epsilon}_y}\epsilon^2
        \log\frac{{\tilde \epsilon}_y}{|\epsilon-vk|}}\\[5mm]
      \hspace{7mm}
        \displaystyle{-\frac{Z^5{\tilde U}^3}{96\pi^3{\tilde \epsilon}_x^3
	{\tilde \epsilon}_y^{3/2}}\frac{\epsilon^3}
        {(\epsilon-vk)^{1/2}}}&\mbox{$(\epsilon>vk)$}\\[5mm]
      \displaystyle{c_1\epsilon+\frac{iZ^5{\tilde U}^3{\tilde K}_0}
	{8\pi^3{\tilde \epsilon}_x^2{\tilde \epsilon}_y}\epsilon^2
        \log\frac{{\tilde \epsilon}_y}{|\epsilon-vk|}}\\[5mm]
      \hspace{7mm}
        \displaystyle{+\frac{iZ^5{\tilde U}^3}{96\pi^3{\tilde \epsilon}_x^3
	{\tilde \epsilon}_y^{3/2}}\frac{\epsilon^3}
        {(vk-\epsilon)^{1/2}}}&\mbox{$(\epsilon<vk)$},
    \end{array}\right.
 \label{eq:pp_res}
\end{eqnarray}
where
\begin{equation}
  c_1\equiv-\frac{U^3}{4\pi^3}\left.\int d^2q
      \left[K^R({\mib q},\omega)\right]^2\,
      \iIm G_q^R(\omega)\right|_{\omega=0},
\end{equation}
and ${\tilde K}_0=K_0\Lambda^{-2}$.
The effect of the first term in
eq.\ (\ref{eq:pp_res}), $c_1\epsilon$, can be also
absorbed into the renormalizations of the constants $v$, $A$ and $Z$.
\def\FIGSCALE{4cm}
\begin{figure}[t]
  \begin{center}
    \begin{tabular}[t]{c}
      \subfigure[]{\epsfxsize=\FIGSCALE\epsfbox{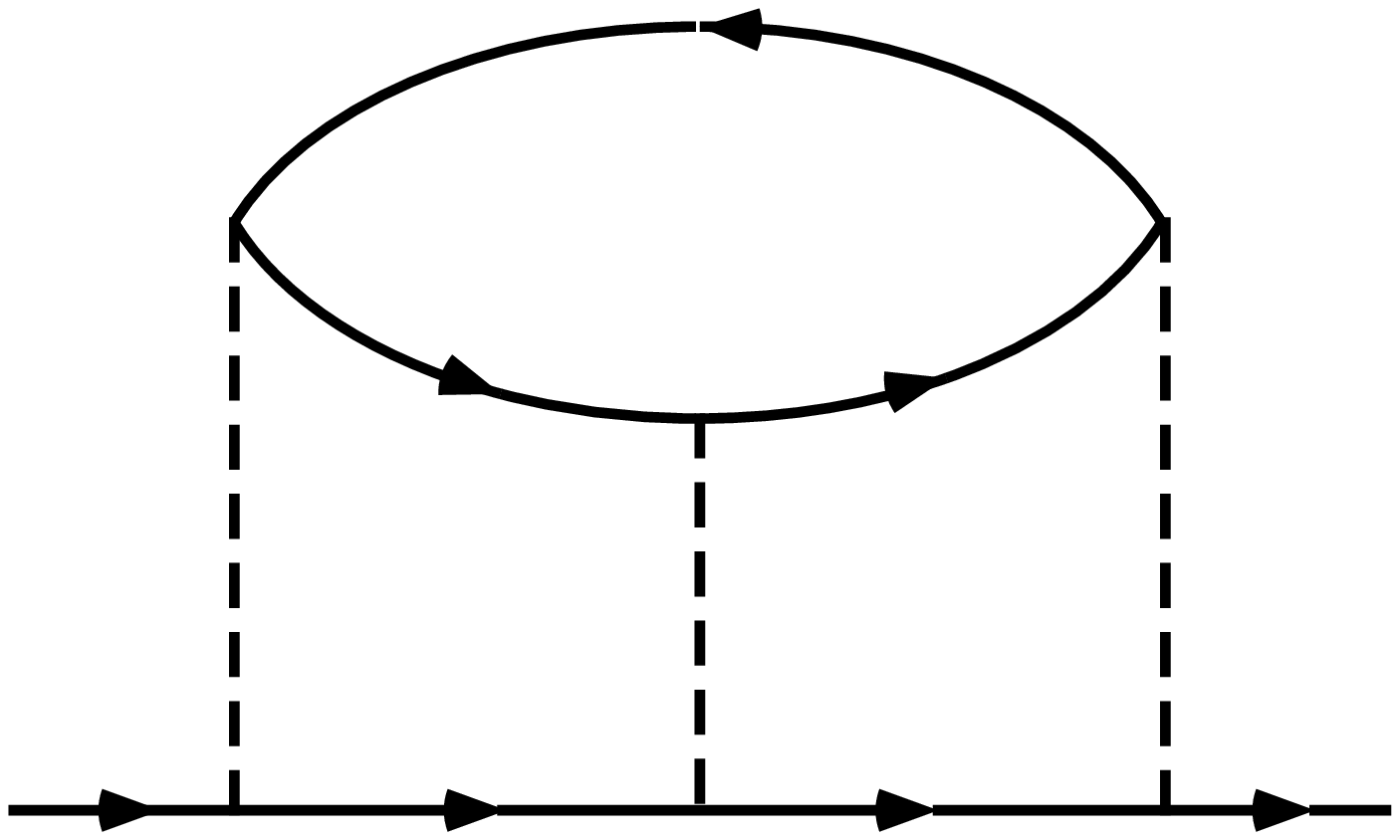}}
      \subfigure[]{\epsfxsize=\FIGSCALE\epsfbox{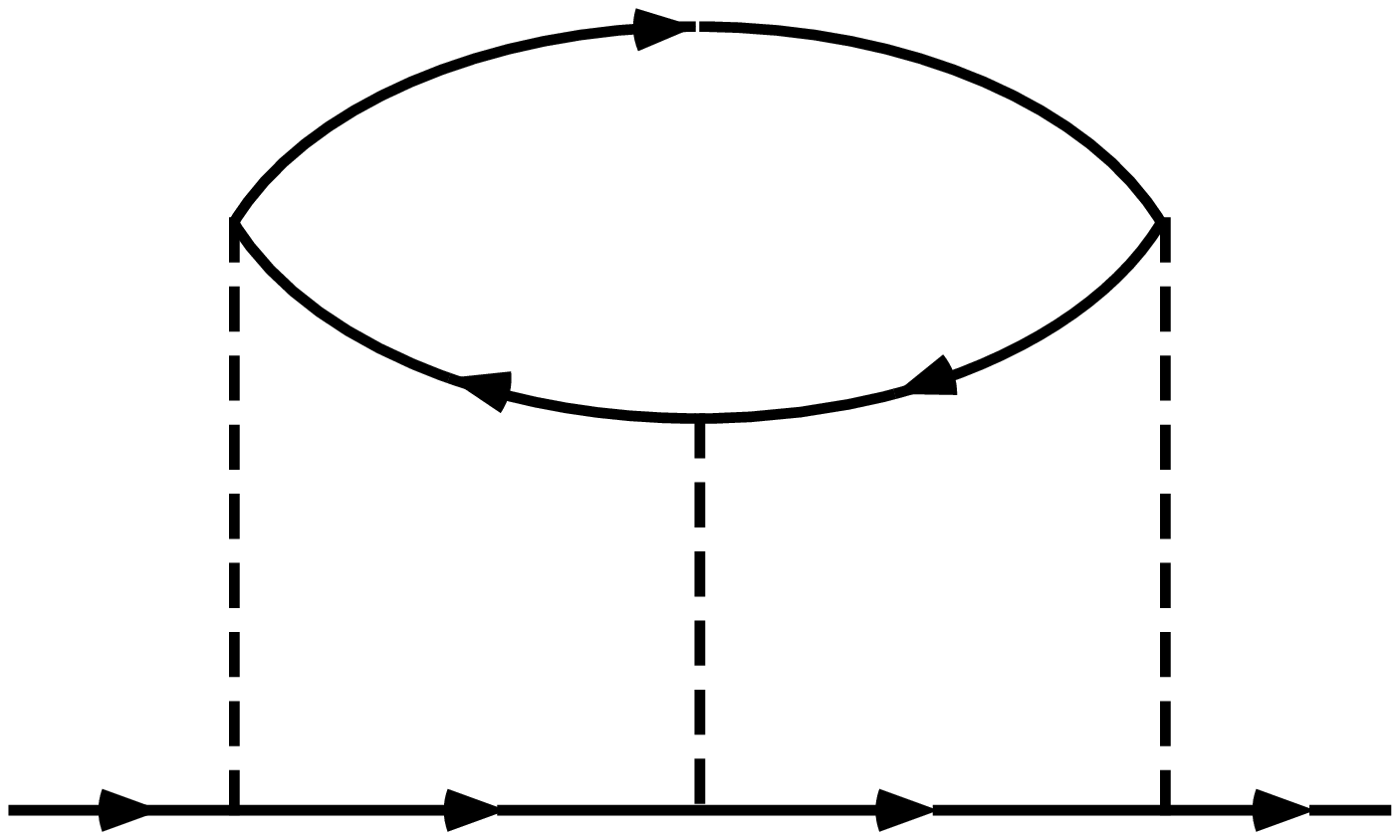}}
    \end{tabular}
  \end{center}
  \caption{Third order self-energy diagrams, which are assumed to cancel out
           in the framework of the loop-cancellation, but do not cancel in
           the present two-dimensional model. (a) Particle-particle
           process. (b) Particle-hole process.}
  \label{fig:3rd}
\end{figure}

Secondly, we consider the contribution of the diagram fig.\ \ref{fig:3rd}(b),
which is denoted as $\Sigma_{ph}^R({\mib k},\epsilon)$ and
expressed as
\begin{eqnarray}
  \lefteqn{\hspace{-6mm}\Sigma_{ph}^R({\mib k},\epsilon)=
    U^3\int_{q\omega}
    \left[\sign(\omega)\,\iIm\left[
    \chi^R({\mib q},\omega)\right]^2\,G_{q+k}^R(\omega+\epsilon)\right.}
    \nonumber\\
  & &\hspace{7mm}\left.+
    \sign(\omega+\epsilon)\,
    \left[\chi^A({\mib q},\omega)\right]^2\,\iIm
    G_{q+k}^R(\omega+\epsilon)\right].
\end{eqnarray}
Here $\chi^{R}({\mib q},\omega)$ is the particle-hole correlation function
defined as
\begin{eqnarray}
  \chi^{R}({\mib q},\omega)&=&\int_{kx}\left[
    \sign(x)\,G_{q+k}^R(\omega+x)\,\iIm G_k^R(x)\right.\nonumber\\
  & &\hspace{-4mm}\left.
    +\sign(\omega+x)\,G_k^A(x)\,\iIm G_{q+k}^R(\omega+x)\right].
\end{eqnarray}
Substituting the energy dispersion given by eq.\ (\ref{eq:disp}),
we obtain the expressions for $\chi^{R}({\mib q},\omega)$ 
in case $|\omega|, |vq_x|, |Aq_y^2|\ll {\tilde \epsilon}_x,
{\tilde \epsilon}_y$ as
\begin{equation}
  \chi^{R}({\mib q},\omega)
  =\left\{\begin{array}{l}
    \displaystyle{\chi_0-\frac{Z^2\omega}{4\pi^2vAq_y}\log\left(
      \frac{A\Lambda q_y-\omega+vq_x}{-A\Lambda q_y-\omega+vq_x} 
      \right)}\\[5mm]
    \hspace{3cm}\mbox{$(|\omega-vq_x|>A\Lambda|q_y|)$}\\[5mm]
    \displaystyle{\chi_0-\frac{Z^2\omega}{4\pi^2vAq_y}\left[\log\left(
      \frac{A\Lambda q_y-\omega+vq_x}{A\Lambda
        q_y+\omega-vq_x}\right)\right.}\\[5mm]
    \displaystyle{\hspace{7mm}+i\pi\,\sign[q_y]\Bigl]}\hspace{5mm}
    \mbox{$(|\omega-vq_x|<A\Lambda|q_y|)$.}
  \end{array}\right.
  \label{eq:ph_cor}
\end{equation}
Here $\chi_0$ is defined as
\begin{equation}
  \chi_0\equiv\lim_{q\rightarrow 0}\lim_{\omega\rightarrow 0}
  \chi^{R}({\mib q},\omega)=-\frac{Z^2\Lambda}{2\pi^2v},
\end{equation}
which is proportional to the density of state at the Fermi energy.
In the same way as in case of the evaluation of
$\Sigma_{pp}^R({\mib k},\epsilon)$,
we obtain the singular part of $\Sigma_{ph}^R({\mib k},\epsilon)$ by
\begin{eqnarray}
  \lefteqn{\hspace{-4mm}{\Sigma_{ph}^R}^\prime({\mib k},\epsilon)}\nonumber\\
  & &\hspace{-3mm}=U^3\!\!\int\!\! \frac{d^2q}{(2\pi)^2}
    \int_{-\epsilon}^{0}\frac{d\omega}{\pi}\,
    \left[\chi^A({\mib q},\omega)\right]^2\,\iIm G_{q+k}^R(\omega+\epsilon),
  \label{eq:ph_self}
\end{eqnarray}
which is evaluated from eqs.\ (\ref{eq:ph_cor}) and (\ref{eq:ph_self})
in case of $\epsilon>vk$, for example, as follows;
\begin{eqnarray}
  \lefteqn{{\Sigma_{ph}^R}^\prime({\mib k},\epsilon)
   \simeq c_2\epsilon}\nonumber\\
  & &-\frac{U^3Z^5\chi_0\epsilon^2}{16\pi^4v^2A}\left\{
    \int_0^{\frac{\epsilon-vk}{A\Lambda}}\!\!\!dq_y\frac{1}{q_y}
    \left[\log
    \frac{A\Lambda q_y+(\epsilon-vk)}{-A\Lambda q_y+(\epsilon-vk)}\right]
    \right.\nonumber\\
  & &\hspace{0.8cm}\left.
    +\int_{\frac{\epsilon-vk}{A\Lambda}}^{2\Lambda}dq_y\frac{1}{q_y}
    \left[\log
    \frac{A\Lambda q_y+(\epsilon-vk)}{A\Lambda q_y-(\epsilon-vk)}-i\pi\right]
    \right\}\nonumber\\
  & &-\frac{U^3Z^5\epsilon^3}{96\pi^6v^3A^2}\left\{
    \int_0^{\frac{\epsilon-vk}{A\Lambda}}\!\!\!dq_y\frac{1}{q_y^2}
    \left[\log
    \frac{A\Lambda q_y+(\epsilon-vk)}{-A\Lambda q_y+(\epsilon-vk)}\right]^2
    \right.\nonumber\\
  & &\hspace{0.8cm}\left.
    +\int_{\frac{\epsilon-vk}{A\Lambda}}^{2\Lambda}dq_y\frac{1}{q_y^2}
    \left[\log
    \frac{A\Lambda q_y+(\epsilon-vk)}{A\Lambda q_y-(\epsilon-vk)}-i\pi\right]^2
    \right\}\nonumber\\
  & &\simeq
    c_2\epsilon+\frac{iU^3Z^5\chi_0}{8\pi^3v^2A}\epsilon^2\log
    \frac{A\Lambda^2}{|\epsilon-vk|}\nonumber\\
  & &\hspace{2.5cm}+\frac{i(\log2)U^3Z^5\Lambda}{24\pi^5v^3A}
    \frac{\epsilon^3}{\epsilon-vk},
\end{eqnarray}
where
\begin{equation}
  c_2\equiv\frac{U^3}{4\pi^3}\left.\int d^2q
      \left[\chi^R({\mib q},\omega)\right]^2\,
      \iIm G_q^R(\omega)\right|_{\omega=0}.
\end{equation}
Evaluating ${\Sigma_{ph}^R}^\prime({\mib k},\epsilon)$ 
for $\epsilon<vk$ in the same way, we obtain the
final result as
\begin{eqnarray}
  {\Sigma_{ph}^R}^\prime({\mib k},\epsilon)&\simeq&
    c_2\epsilon+\frac{iZ^5{\tilde U}^3{\tilde \chi}_0}
    {8\pi^3{\tilde \epsilon}_x^2{\tilde \epsilon}_y}\epsilon^2\log
    \frac{{\tilde \epsilon}_y}{|\epsilon-vk|}\nonumber\\
  & &\hspace{0.7cm}+\frac{i(\log2)Z^5{\tilde U}^3}
    {24\pi^5{\tilde \epsilon}_x^3{\tilde \epsilon}_y}
    \frac{\epsilon^3}{\epsilon-vk},
  \label{eq:ph_res}
\end{eqnarray}
where ${\tilde \chi}_0=\chi_0 \Lambda^{-2}$.

As for second-order term, we can obtain the singular part
in the same way as follows;
\begin{equation}
  {\Sigma_{2nd}^R}^\prime({\mib k},\epsilon)\simeq
    c_3\epsilon-\frac{iU^2Z^3\epsilon^2}{16\pi^3v^2A}
    \log\frac{A\Lambda^2}{|\epsilon-vk|},
  \label{eq:pp_2nd}
\end{equation}
where
\begin{equation}
  c_3\equiv\frac{U^2}{4\pi^3}\left.\int d^2q
      K^R({\mib q},\omega)
      \iIm G_q^R(\omega)\right|_{\omega=0}.
\end{equation}

From eqs.\ (\ref{eq:pp_res}) and (\ref{eq:ph_res}),
we have to notice that there exists a region where
$\iIm \Sigma^R(k,\epsilon)$ has a positive value, although
$\iIm \Sigma^R(k,\epsilon)$ has to be negative-definite.
This is due to the reason that we have considered only up to the
third order.
Eqs.\ (\ref{eq:pp_res}), (\ref{eq:ph_res}) and (\ref{eq:pp_2nd})
implies the possibility
that the spectral-weight $\pi^{-1}|\iIm G^R(k,\epsilon)|$
has a two-peak structure in the vicinity of $\epsilon=\xi_k$
due to the divergence of the self-energy at $\epsilon=\xi_k$.
The singular part of the self-energy is relevant only in the vicinity of
$\epsilon=\xi_k$, and we can define the width $\Delta(k)$
of the structure of the spectral-weight in this vicinity as
$\Delta(k)=|\Sigma^R(k,vk+\Delta(k))|$.
Evaluating $\Delta(k)$ for ${\Sigma_{pp}^R}^\prime$,
${\Sigma_{ph}^R}^\prime$ and ${\Sigma_{2nd}^R}^\prime$,
which we denote $\Delta_{pp}(k)$, $\Delta_{ph}(k)$ and $\Delta_{2nd}(k)$,
respectively, we obtain
$\Delta_{pp}(k)\simeq(Z^{10/3}\tilde{U}^2
{\tilde\epsilon}_x^{-2}{\tilde\epsilon}_y^{-1})|vk|^2$,
$\Delta_{ph}(k)\simeq(Z^{5/2}\tilde{U}^{3/2}
{\tilde\epsilon}_x^{-3/2}{\tilde\epsilon}_y^{-1/2})|vk|^{3/2}$ and
$\Delta_{2nd}(k)\simeq(Z^3\tilde{U}^2
{\tilde\epsilon}_x^{-2}{\tilde\epsilon}_y^{-1})|vk|^2
\log({\tilde\epsilon}_y/|vk|)$.
Noticing the exponents of $|vk|$ in the expressions for $\Delta(k)$,
$\Delta_{ph}(k)$ is dominant for small $|k|$
(here the origin of $k$ is taken on the Fermi surface)
reflecting the fact that ${\Sigma_{ph}^R}^\prime$ has
a stronger divergence at $\epsilon=vk$ than ${\Sigma_{pp}^R}^\prime$,
which indicates that
the particle-hole process (fig.\ 3(b)) has a tendency to
enhance the splitting of the spectral-weight in
the vicinity of $\epsilon=\xi_k$. We can also see
$\Delta(k)\ll|\xi_k|$ for small $|\xi_k|$,
which indicates that the quasi-particle picture is valid
for sufficiently low-energy.
$\Delta(k)$ grows as $|\xi_k|$ becomes larger, which
leads to the crossover from Fermi to non-Fermi liquid behavior
as the momentum $k$ becomes away from the Fermi momentum.
This crossover-energy $\Delta^c$ can be defined as
$\Delta(k)$ which satisfies $\Delta(k)=|\xi_k|$.
Above expressions for $\Delta(k)$ is not applicable for
$|vk|\gg {\tilde\epsilon}_x, {\tilde\epsilon}_y$,
however, we can see the tendency of the interaction dependence
of this crossover energy by assuming above expressions
for $\Delta(k)$ for
$|\xi_k|\gg {\tilde\epsilon}_x, {\tilde\epsilon}_y$.
The explicit expression for $\Delta(k)$ obtained by this way is not valid,
but can be considered to reflect the speed of growth of the splitting
of the spectral-weight as $|\xi_k|$ becomes larger.
Denoting $\Delta^c$ evaluated from $\Delta_{pp}(k)$,
$\Delta_{ph}(k)$ and $\Delta_{2nd}(k)$ as
$\Delta^c_{pp}$, $\Delta^c_{ph}$ and $\Delta^c_{2nd}$,
we obtain $\Delta^c_{pp}\simeq Z^{-10/3}\tilde{U}^{-2}
{\tilde\epsilon}_x^2{\tilde\epsilon}_y$,
$\Delta^c_{ph}\simeq Z^{-5}\tilde{U}^{-3}{\tilde\epsilon}_x^3
{\tilde\epsilon}_y$ and
$\Delta^c_{2nd}\simeq Z^{-3}\tilde{U}^{-2}
{\tilde\epsilon}_x^2{\tilde\epsilon}_y\log(Z^{3/2}U/{\tilde\epsilon}_x)$,
which indicates $\Delta^c\gg{\tilde\epsilon}_x, {\tilde\epsilon}_y$
for small $U$.
Although this represents that the crossover 
arises only at higher energy compared to band-width
${\tilde\epsilon}_x$ and ${\tilde\epsilon}_y$ in weak-coupling case,
we can see a tendency that the crossover-energy decreases
as the interaction $U$ becomes stronger.
From above consideration within the scope of weak-coupling,
a possibility is expected
that the crossover energy becomes smaller than band-width for
strong-coupling. This possibility is interesting in
relation to the non-Fermi liquid behavior of the High-$T_c$ cuprates
in low-doping region.

Furthermore, we can consider above results from the viewpoint of
loop-cancellation;\cite{bib:Metrev}
the third-order diagrams in fig.\ \ref{fig:3rd}
are assumed to cancel out in the framework of loop-cancellation.
The loop-cancellation is exact for TL model, while
in case of higher dimensions, the cancellation is not complete
even for forward scattering process.
Metzner et al.\ made an elementary estimation by power
counting about the residual term of the cancellation for
loop-diagrams,\cite{bib:loop}
while the contribution of this residual term of loop-diagrams to
the self-energy is not precisely estimated.
From eqs.\ (\ref{eq:pp_res}) and (\ref{eq:ph_res}), we can see that
the singularities of the contributions from the two third-order
diagrams do not cancel. This implies that the loop-cancellation is
insufficient, and the residual contribution
is relevant in the vicinity of $\epsilon=vk$.
This can be considered as follows;
although for completely flat Fermi surface,
the exact loop-cancellation holds
even in two-dimension, there is a warping of the
Fermi surface characterized by the energy scale ${\tilde\epsilon}_y$
in general.
In case of $|\epsilon|,|vk|\gg {\tilde\epsilon}_y$,
the loop-cancellation is a good 
approximation, and the self-energy can be expanded by
small parameters ${\tilde\epsilon}_y/|\epsilon|,|vk|$.
If we consider the low-energy
limit, however, the effect of warping becomes crucially important;
${\tilde\epsilon}_y/|\epsilon|,|vk|$ is no more a small parameter
in case of $|\epsilon|,|vk|\ll {\tilde\epsilon}_y$. This implies
that the loop-cancellation is not suitable as a starting point
to consider the low-energy excitation
in two-dimension even for forward scattering model.
Especially, the effect of the residual term is noticeable
in the vicinity of $\epsilon=vk$.

The author would like to thank to H. Fukuyama for valuable
discussions. Thanks are also due to K. Miyake, O. Narikiyo and
H. Maebashi for valuable comments. The author also acknowledges
JSPS Research Fellowship for Young Scientists.

\end{document}